# Ionization and excitation of the excited hydrogen atom in strong circularly polarized laser fields

Jarosław H. Bauer[1,*], Francisca Mota-Furtado[2], Patrick F. O'Mahony[2], Bernard Piraux[3], and Krzysztof Warda[4]

[1]*Katedra Fizyki Teoretycznej Uniwersytetu Łódzkiego, Ul. Pomorska 149/153, PL-90-236 Łódź, Poland*

[2]*Department of Mathematics, Royal Holloway, University of London, Egham, Surrey TW20 0EX, United Kingdom*

[3]*Institute of Condensed Matter and Nanosciences, Université Catholique de Louvain, chemin du Cyclotron, 2 bte L7.01.07, 1348 Louvain-la-Neuve, Belgium*

[4]*Katedra Fizyki Ciała Stałego Uniwersytetu Łódzkiego, Ul. Pomorska 149/153, PL-90-236 Łódź, Poland*

In the recent work of Herath *et al.* [Phys. Rev. Lett. **109**, 043004 (2012)] the first experimental observation of a dependence of strong-field ionization rate on the sign of the magnetic quantum number $m$ (of the initial bound state $(n,l,m)$) was reported. The experiment with nearly circularly polarized light could not distinguish which sign of $m$ favors faster ionization. We perform *ab initio* calculations for the hydrogen atom initially in one of the four bound sub states with the principal quantum number $n=2$ and irradiated by a short circularly polarized laser pulse of $800\,nm$. In the intensity range of $10^{12}-10^{13}\,W/cm^2$ excited bound states play a very important role, but also up to some $10^{15}\,W/cm^2$ they can not be neglected in a full description of the laser-atom interaction. We explore the region that with increasing intensity switches from multiphoton to over the barrrier ionization and we find unlike in tunneling-type theories, that the ratio of ionization rates for electrons initially counter-rotating and co-rotating (with respect to the laser field) may be higher or lower than one.

---

*bauer@uni.lodz.pl

# I. INTRODUCTION

The progress of strong-field laser-atom physics has been caused and significantly motivated by the first experimental observations of above threshold ionization (ATI) [1-3] of atoms more than thirty years ago. ATI can be usually understood as a process of multiphoton absorption, when the atom (or ion or molecule) absorbs more photons from an electromagnetic (laser) field than the minimum number required to overcome the ionization threshold. Alternatively, for sufficiently intense fields and sufficiently low frequencies of the laser, the process may be understood as a tunneling through a suppressed barrier of the Coulomb potential [4,5]. For even stronger laser fields a barrier-suppression ionization (BSI) (called also above-barrier or over-the-barrier ionization) takes place [4-6], because the total resulting potential (Coulomb and laser) cannot hold the electron above the (negative) initial-state energy. In the present work we obtain results pertaining to the first and third regimes of the laser field parameters.

Usually, the Keldysh parameter $\gamma$ [7] is used to distinguish between multiphoton and tunneling ionization.

$$\gamma = \frac{\omega\sqrt{2E_B}}{F} = \frac{Z\omega}{nF} , \qquad (1)$$

where $\omega$ is the laser frequency, $F$ - the amplitude of the laser field, and $E_B = Z^2/(2n^2)$ - the binding energy of the atom (of the nuclear charge $Z$), initially in the state described by the well-known $(n,l,m)$ quantum numbers (without spin), in the nonrelativistic approach . (We use atomic units (a.u.) in the present work, with the peak laser intensity given by $I = 2F^2$ for circularly polarized (CP) radiation. $1\ a.u. = 3.51 \cdot 10^{16} W/cm^2$.) Multiphoton ionization dominates when $\gamma > 1$, and tunneling ionization prevails when $\gamma < 1$. The latter statement is valid only after further specification. Another important parameter is the barrier-suppression field strength (see, for example, [4-6])

$$F_{BSI} = \frac{E_B^2}{4Z} = \frac{Z^3}{16n^4} . \qquad (2)$$

Let us note that when, approximately,

$$F \leq F_{BSI} \tag{3}$$

(see [8]), the Keldysh parameter measures the ratio of the so-called barrier penetration time to half the laser cycle [7,9]. If the laser frequency is too high, the ionized electron does not have enough time to tunnel out before the electric field vector changes (for the CP field of constant intensity only its direction changes). Hence, tunneling is impossible in this case. Sometimes $\gamma$ is called the adiabaticity parameter [5,9-10]. If $F \leq F_{BSI}$ and $\gamma \ll 1$, the so-called adiabatic tunneling becomes the dominant mechanism of strong-field ionization.

Typically, in review papers describing phenomena in intense laser fields (see, for instance, [5,10-12]), relatively less attention is paid to atoms in bound states described by principal quantum numbers $n \geq 2$ and CP fields. Indeed, only recently the first experimental observation of the dependence of the strong-field ionization rate on the sign of magnetic quantum number $m$ was reported. In the experiment of Herath *et al.* [13], with argon atoms in nearly CP laser fields, the pump-probe technique was applied to valence ($3p$) electrons. By measuring the dication ($Ar^{2+}$) yield in sequential double ionization of argon, different ionization rates for electrons co-rotating and counter-rotating with respect to the electric field vector of the laser, have been experimentally confirmed. However, unlike as stated by the authors in Ref. [13], not only one theoretical work, namely [14], discussed such effects before. (In both works [13,14] the standard $Ti:sapphire$ 800 $nm$ laser radiation was used.) The list of papers showing analogous results is much longer, although certainly not complete [15-21] (some of them were published before or soon after Ref. [13]). In some of these earlier works the time-dependent Schrödinger equation (TDSE) (for hydrogenic atoms in CP laser fields) was solved numerically (see, for example, [15-17]), and for the same laser pulse, different ionization probabilities and photoelectron energy spectra were obtained, depending on $(n,l,m)$ quantum numbers. On the other hand, the famous Keldysh strong-field approximation in the length gauge [7] can be generalized to CP laser fields. Using this approach, different ionization rates (total or differential) were obtained for the hydrogen atom in initial states with different $(n=2,l,m)$ quantum numbers [18-21]. It is worth noting that only the length gauge version of the $S$-matrix theory gives different ionization rates and photoelectron energy spectra for the states $(2,1,-1)$ and $(2,1,1)$ [19,21]. It is worth noting also that standard quasi-static or adiabatic tunneling theories predict that these rates and spectra should depend only on the absolute value of the magnetic quantum number $m$, but not on its sign [10].

The experiment of Herath *et al.* [13] was interpreted by Barth and Smirnova [14,22]. They used the analytic theory of Perelomov, Popov and Terent'ev [23] for short range potentials which covers in particular the non-adiabatic tunneling region where $F \leq F_{BSI}$ and $\gamma \approx 1$ [24]. They later included corrections due to the long range Coulomb field [25]. As a result, good quantitative agreement between theory and experiment has been found [26]. The experiment, however, could not distinguish which sign of $m$ favors faster ionization. But, according to the predictions of Refs. [14,22], counter-rotating electrons (in the initial bound state) should ionize much faster than co-rotating electrons. This is true in the whole range for $\gamma > 0$. In the adiabatic limit $\gamma \ll 1$ both ionization rates become nearly equal (cf. Eq. (10) in Ref. [14]). After Ref. [14], the ratio (counter-rotating to co-rotating) of ionization rates grows monotonically from $1 + 4\gamma/3$ to $(2\ln\gamma)^2$, on increasing the Keldysh parameter $\gamma$ from $\gamma \ll 1$ to $\gamma \gg 1$. Thus, it follows from Eq. (1) that lower laser intensities should always amplify the measurable effect (namely, the ratio of ionization yields). We have verified that in both Refs. [13] and [14] the condition given by Eq. (3) is satisfied. Indeed, for *Ar* atoms $E_B = 0.58$ *a.u.*, and for $Ar^+$ ions $E_B = 1.0$ *a.u.*. Then, using the effective principal quantum numbers $n^* = Z/\sqrt{2E_B}$ [10] (with $Z = 1$ for *Ar* and $Z = 2$ for $Ar^+$) in Eq. (2), one obtains $F_{BSI} = 0.084$ *a.u.* for *Ar* atoms and $F_{BSI} = 0.13$ *a.u.* for $Ar^+$ ions. This corresponds to $I_{BSI} = 4.9 \cdot 10^{14} W/cm^2$ and $I_{BSI} = 1.2 \cdot 10^{15} W/cm^2$, respectively. Herath *et al.* [13] did their experiment with $I = 9 \cdot 10^{13} W/cm^2$ and $I = 1.4 \cdot 10^{14} W/cm^2$, respectively. These intensities correspond to (cf. Eq. (1); see also Ref. [26]) $\gamma = 1.7$ for *Ar* and $\gamma = 1.8$ for $Ar^+$. In Fig. 2 of Ref. [14] ionization rates are calculated for electrons ionized from krypton (ground $4p$ state), where $E_B = 0.52$ *a.u.*. In this figure the intensity range $2.0 \cdot 10^{13} W/cm^2 \leq I \leq 2.0 \cdot 10^{14} W/cm^2$ is presented. This corresponds to $3.4 \geq \gamma \geq 1.1$. Since $Z = 1$ for a single ionization of *Kr* atoms, one obtains $I_{BSI} = 1.6 \cdot 10^{14} W/cm^2$, which means that this is (almost) the range of nonadiabatic tunneling.

In this paper we consider a different regime to that considered by Barth and Smirnova where our initial state is mainly over the barrier at the peak of the electric field pulse (BSI). For the four $n=2$ initial states of hydrogen we calculate the total probability that the atom is left in an excited state, as well as the ionization probability, by solving the TDSE numerically for a right circularly polarized pulse. We find that unlike in the tunneling regime in [14] where excited atomic states are considered to play a negligible role, they play a crucial role in the BSI region and we find significant population of excited states even for quite high field intensities. In addition we find that for low intensities the ionization yield for the co-rotating initial state (2,1,1) is higher than for the counter-rotating initial state (2,1,-1) but this behaviour switches over as we increase the intensity.

## II. THEORY

Conventionally, the nonadiabatic tunneling regime could be specified by its lower and upper limits, namely $I_{min} = 4\omega^2 E_B$ (when $\gamma = 1$) and $I_{max} = I_{BSI} = 2F_{BSI}^2$, respectively. If $I_{min} > I_{max}$, approximately, then there is no nonadiabatic tunneling regime. This is equivalent to the following condition:

$$\omega > \omega_{\lim} \equiv \frac{F_{BSI}}{\sqrt{2E_B}} = \frac{1}{Z}\sqrt{\frac{E_B^3}{2^5}} \ . \tag{4}$$

If Eq. (4) is satisfied, tunneling-type models cease to suffice. In terms of the Keldysh parameter $\gamma$, increasing the peak intensity, one can go directly from the domain of multiphoton absorbtion to the domain of BSI for specific laser field parameters. In much of the intensity range we shall describe the initial state lies above the barrier created at the instantaneous peak intensity of the laser. Scrinzi et al [6] found for linearly polarised pulses that for such regions where BSI dominates, a relevant parameter is a critical field strength $F_c$, which is independent of laser frequency, such that for fields $F > F_c$ the adiabatic approximation is applicable.

To investigate the BSI region we perform accurate *ab initio* calculations (see, for example, Ref. [27] for CP fields). To this end, we have solved numerically the TDSE with a method similar to the one described in Refs [15,16] where benchmark results were presented for atomic hydrogen exposed to CP fields. Very briefly, we use the velocity gauge (with well-known advantages [28]) and the dipole approximation and assume the following form of the vector potential:

$$\vec{A}(t) = A_0 f(t)[-\vec{e}_x \sin(\omega t) + \vec{e}_y \cos(\omega t)]. \tag{5}$$

In this formula, $A_0 = F_0/\omega$ is the amplitude of the vector potential, $\omega$ is the frequency of the field, $\vec{e}_x$ and $\vec{e}_y$ are the unit vectors in the x and y directions, respectively, and $f(t)$ is a slowly varying pulse envelope. In our calculations, $f(t)$ has a sine-squared form:

$$f(t) = (\sin(\pi t/t_d))^2, \tag{6}$$

where $t_d$ is the pulse time duration. If we assume the frequency $\omega$ positive, the above form of the vector potential corresponds to $\sigma^+$ polarization. In order to solve the TDSE, we use a spectral method which consists of expanding the total wave function as follows:

$$\Psi(\vec{r},t) = \sum_{l=0}^{l_{max}} \sum_{m=-l}^{l} \sum_{n=l+1}^{l+1+N} a_{nlm}(t) \frac{1}{r} S_{n,l}^{\kappa}(r) Y_{l,m}(\theta,\varphi). \tag{7}$$

In this expression, $S_{n,l}^{\kappa}(r)$ is a Coulomb-Sturmian function of the electron radial coordinate $r$, $Y_{l,m}(\theta,\varphi)$ is the usual spherical harmonics of the electron angular coordinates and $a_{nlm}(t)$ is a coefficient. The Coulomb-Sturmian functions which are studied in detail in [16] form a discrete and complete set in the space of the $L^2$-integrable functions. These functions depend on the parameter $\kappa$. For any given $\kappa$ the Coulomb-Sturmian functions form a complete set. Since in the previous expansion, we use a finite number $N$ of Coulomb-Sturmian functions per $(l,m)$ pair, the choice of the value of $\kappa$ is relevant. The fact that $\kappa$ behaves as a dilation parameter and that for $\kappa = 1/n$, the corresponding Coulomb-Sturmian function $S_{n,l}^{\kappa}(r)$ coincides with the radial hydrogen bound state wave function $R_{n,l}(r)$, provides a method of controlling, for a given $N$, the number of hydrogen bound states that are correctly reproduced in our basis. This method is discussed in detail in [16] and has been generalized to the case of helium in [29].

The previous discussion clearly indicates that the Coulomb-Sturmian basis is one of the best adapted ones for the case of atomic hydrogen. Substituting the expansion Eq. (7) into the TDSE we get a set of coupled first order time dependent equations for the coefficients $a_{nlm}(t)$. The resulting set of equations is very large for a circularly polarised laser pulse since we have a sum not only over $(n, l)$ but also over $m$. However, both the atomic and the interaction Hamiltonian can be very easily expressed in this basis: firstly, all matrix elements have a very simple analytical expression and second, the total Hamiltonian matrix is very sparse. It is in fact block tridiagonal, the diagonal blocks being tridiagonal and the off-diagonal blocks four-diagonal. The sparsity of all matrices allows one to treat large scale problems while exploiting parallelism. It can be shown that the number of $(l, m)$ pairs included in our calculations is given by $(l_{max} + 1)(l_{max} + 2)/2$. This number fixes the maximum number of processors that can be used in parallel.

The main drawback of the present method concerns the time propagation. In any basis that differs from the atomic basis (in which the matrix associated to the atomic Hamiltonian is diagonal), the system of first order differential equations to solve behaves as a highly stiff system. In [15,16], this difficulty was overcome by using a diagonally implicit fourth order Runge-Kutta method. However, the implicit character of this method requires the solution of many very large systems of algebraic equations at each time step making the computing time excessive and limiting an efficient parallelism to four processors only. Here, we use the explicit one-step algorithm of Arnoldi which is a Krylov subspace method. The way it is used in the present context is explained in detail in [30].

The spectrum, ionization probabilities and excitation probabilities are calculated by projecting onto bound and continuum states of the field free problem, namely hydrogen, at the end of the laser pulse.

### III. RESULTS

In Fig. 1 we show probabilities for ionization, excitation, and remaining in the initial state of the hydrogen atom, namely $(2,1,-1)$, as a function of the peak laser intensity. In Figs. 2, 3, and 4 we show analogous probabilities for the initial states $(2,1,1)$, $(2,1,0)$, and $(2,0,0)$, respectively. All the probabilities are calculated after switching off the laser pulse (therefore, they are gauge invariant). $\tau = 10 \cdot 2\pi / \omega$, i.e. 10 cycles, is the total duration of the laser pulse with a sine squared envelope. $\omega = 0.057$ *a.u.* conforms with *Ti : sapphire* 800 *nm* laser radiation. The condition from Eq. (4) is satisfied in our case (the right-hand side is equal to

0.0078 *a.u.*). In the lowest order of perturbation theory three photons are needed to overcome the ionization threshold at $E_B = 0.125$ *a.u.*. The excitation is understood here as a sum of populations over all bound states except the initial one. Figures 1-4 clearly show that when the peak intensity of the laser field grows, beginning from the perturbative regime, the excitation grows faster than the ionization initially. Indeed, substantial ionization starts at the peak intensity close to the value of $I_{BSI} = 2F_{BSI}^2 = 1.1 \cdot 10^{12} W/cm^2$, but the excitation starts at the peak intensity about 10 times smaller. The dominant mechanism for excitation and ionisation below $I_{BSI}$ is clearly multiphoton transitions. Above this intensity, eventhough the initial state lies above the barrier for the peak intensity of the pulse, we expect multiphoton transitions to dominate for these large values for $\gamma$. We note that the excitation probability may be as high as $0.4 - 0.8$ at its maximum. The latter is achieved for similar peak intensities (for all initial states with $n = 2$), lying in the range $10^{12} - 10^{13} W/cm^2$, corresponding to a quiver radius of the order of $2 - 3$ *a.u.*, depending on the initial state. Visibly higher intensity (about $6 \cdot 10^{12} W/cm^2$) is needed for the state $(2,1,-1)$, to achieve the highest possible excitation. The hydrogen atom in the states $(2,1,1)$ and $(2,1,0)$ may be more excited (by adjusting properly the peak intensity) than for the other two states. When the peak intensity increases further, the excitation decreases, achieving finally a level of some $0.1 - 0.2$. At the same time, the ionization grows up to the level of $0.8 - 0.9$. Figures 1-4 show that above $10^{15} W/cm^2$ the ionization is much bigger than the excitation, and the latter is much bigger than the population remaining in the initial state, where much less than 1% of atoms survive. Significant excitations of Mg atoms in CP fields were observed in *ab initio* calculations [31]. The authors of this work used a rescattering model to explain why a large number of electrons might return to the parent ion and finish their trajectories in states of negative energies, namely Rydberg states. However they were in a regime where the initial state was below the barrier and tunneling type theories should be applicable.

Above $I \approx 6 \cdot 10^{13} W/cm^2$ excitations remain nearly constant or may even decrease slightly (cf. Figs. 1 and 2). As we increase the intensity multiphoton excitation/ionization should give way to over the barrier ionization and indeed it is suprising that there is significant excitation remaining for high intensities. This may be coming from the weaker part of the laser pulse (near its end). For the initial state $(2,0,0)$, see Fig. 4, there is a second peak in the excitation probability at about $I = 10^{13}$ $W/cm^2$ which coincides with a dip in the ionization probaility. We speculate that this may be mediated by an intermediate resonance in the multiphoton excitation.

In Fig. 5 we show, how fast each of the four above-mentioned initial states ionize with increasing peak intensity. (We compare exactly the same curves, which have been already presented in Figs. 1-4 and marked as "ionization".) For very low peak intensities, roughly up to $9 \cdot 10^{11} W/cm^2$, ionization of the state $(2,1,1)$ is the fastest. Starting from this value, and up to about $10^{15} W/cm^2$ the fastest ionization occurs for the state $(2,0,0)$. For the other three states a dependence of the ionization probability on the peak intensity is rather monotonic and usually increasing. But for the state $(2,0,0)$ there is a dip near $10^{13} W/cm^2$, which corresponds to a hump in the excitation curve (cf. Fig. 4). Only after exceeding the peak intensity of about $5 \cdot 10^{12} W/cm^2$ the ionization probability of the state $(2,1,0)$ becomes the smallest one among the four. Near the value of $10^{13} W/cm^2$ the ionization probability of the state $(2,1,-1)$ exceeds the ionization probability of the state $(2,1,1)$. Thus, the ionization rate of $(2,1,-1)$ must be higher than the ionization rate of $(2,1,1)$ above $10^{13} W/cm^2$. This situation persists up to the highest peak intensities shown in this plot. For low intensities where the multiphoton process dominates it is well known that the ionization rate for $(2,1,1)$ should be larger than that for $(2,1,-1)$ [14]. For increasing intensity, as over the barrier ionization dominates, we expect from the work of Scrinzi et al [6] for linear polarisation, that we should approach the adiabatic limit. This would explain the transition seen near $10^{13} W/cm^2$ where ionization from the $(2,1,-1)$ becomes favoured as seen by Barth and Smirnova [14,22].

In Fig. 6 we show, how fast each of the initial states deplete with increasing peak intensity. (We compare exactly the same curves, which have been already presented in Figs. 1-4 and marked as "initial state".) Roughly up to $3 \cdot 10^{12} W/cm^2$ the fastest depletion of the initial state (i.e. the initial-state probability is the smallest) occurs for the state $(2,1,1)$. Near $10^{13} W/cm^2$ there is a small maximum in this curve, where the probability reaches about $0.1$. Similar small maxima, for slightly different peak intensities, occur for the states $(2,1,0)$ and $(2,0,0)$, but not for $(2,1,-1)$. Above $10^{15} W/cm^2$ the initial-state population is below $0.01$ in all four cases. Up to about $7 \cdot 10^{12} W/cm^2$ the state $(2,1,1)$ depletes faster (initially much faster) with increasing the peak intensity than the state $(2,1,-1)$. Above $7 \cdot 10^{12} W/cm^2$ the situation reverses. Regarding the states $(2,1,-1)$ and $(2,1,1)$, it follows from a comparison of Figs. 5 and 6 that the ratios of respective ionization and depletion rates may be higher or lower than one, depending on the intensity range. Moreover, transitions from one range to the other occur for slightly different peak intensities for the ionization and for the depletion. This is connected with the fact that the state $(2,1,1)$ undergoes a bigger excitation than the state $(2,1,-1)$. Furthermore, lower intensity is needed in the former case to achieve the peak of the excitation probability (cf. Figs. 1 and 2).

## IV. SUMMARY AND OUTLOOK

In conclusion, we have investigated ionization and excitation of the $(n=2,l,m)$ hydrogen atom by the CP 800 *nm* 10-cycle laser pulse in a broad range of peak intensities, covering almost five orders of magnitude. Laser light of fixed (positive) helicity has been employed in our calculations, so the sign of the angular momentum component along the propagation direction axis for the states $(2,1,1)$, (co-rotating electron) and $(2,1,-1)$, (counter-rotating) is the same and the opposite, respectively. We have identified the lower limit $\omega_{lim}$ for the laser frequency (cf. Eq. (4)), above which atoms go directly from multiphoton to the BSI regime by increasing the intensity, without passing through the tunneling regime. We have shown that for the standard laser frequency $\omega = 0.057$ *a.u.* $> \omega_{lim}$, utilized in our calculations, we have found significant probabilities for leaving the atom in an excited state at the end of the pulse even at quite high intensities. The ionization rate behavior as a function of the laser intensity is very different from that of tunneling-type theories. This is not unexpected as we are in very different physical regime. Moreover, it appears that with increasing intensity, very strong excitation of the atom takes place before the strong-field ionization. Therefore, both the ionization rate and the rate of depletion of the initial state have to be taken into account in a theoretical description of the process. Unlike in tunneling-type theories, the ratio of ionization rates for electrons initially counter-rotating and co-rotating (with respect to the laser field) may be higher or lower than one. Initially, with increasing the intensity, co-rotating electrons ionize faster, but then counter-rotating electrons ionize faster (cf. Fig. 5 for the states $(2,1,1)$ and $(2,1,-1)$). The fastest ionization usually takes place for the state $(2,0,0)$, and the slowest one - for the state $(2,1,0)$.

Recently, in Ref. [32], photoelectron angular distributions of excited hydrogenic atoms in intense laser fields have been studied. The authors of this work were using a semianalytical Keldysh theory in a broad range of the Keldysh parameters (both $\gamma < 1$ and $\gamma > 1$) for both circular and linear polarizations and for some initial bound states up to $n = 4$. In the $S$-matrix type calculations, like for instance [14,32], the ionization rate is calculated, the initial-state occupation is assumed to be close to one, and excitations are completely neglected. Of course, depletion effects may be evaluated by using the well-known formula $1 - \exp\left(-\int_0^\tau \Gamma(I(t))dt\right)$ for the laser pulse, where $\Gamma$ denotes the ionization rate. Our work shows that there is always a range of peak intensities (in our case for $\gamma > 1$) where the excitation is much more probable than the ionization. With growing intensity the excitation decreases, but even for very strong fields (above $10^{15} W/cm^2$) it remains significant (cf. Figs. 1-4). This raises questions as to the validity of the Keldysh approach, which neglects excitation, when considering initial states for $n = 2$ or higher in hydrogen for the frequency considerd here.

For the hydrogen $n = 2$ states considered in this paper, to compare numerical results with the predictions of the analytic theory of Barth and Smirnova [14,22] in the nonadiabatic tunneling regime, requires $\omega < \omega_{\lim} = 0.0078$ $a.u.$ in our case (cf. Eq. (4)). Such low frequencies for CP laser fields pose a very challenging problem in numerical calculations which we hope to address.

## ACKNOWLEDGEMENTS

F.M.F. , P.F.O'M and B.P. thank the E.U. COST network through the Action CM1204 "XUV/X-ray light and fast ions for ultrafast chemistry (XLIC) for financing short short term scientific missions. Computational resources have been provided by the supercomputing facilities of the Université Catholique de Louvain and the Consortium des Equipements de Calcul Intensif en Fédération Wallonie Bruxelles (CECI) funded by the Fonds de la Recherche Scientifique de Belgique (F.R.S.-FNRS) under convention 2.5020.11. The present paper has been supported by the University of Łódź.

**FIGURES**

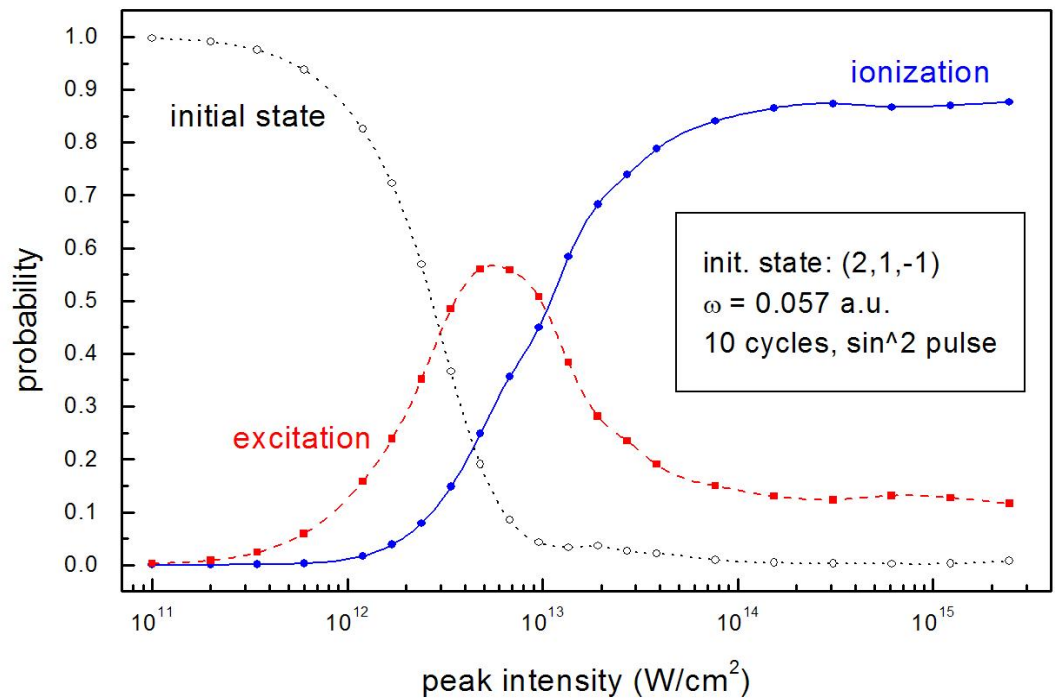

FIG. 1. (Color online) Probability of ionization (blue solid line with solid circles), excitation (red dashed line with solid squares) and remaining in the initial state of the hydrogen atom

(black dotted line with open circles) as a function of the peak laser intensity (after switching off the laser pulse). Each circle or square is a result of a single simulation - numerical solution of the TDSE. Connecting lines are calculated with the help of splines. The initial state is $(2,1,-1)$ (see the text for more detail). The laser light is counter-rotating with respect to the initial state.

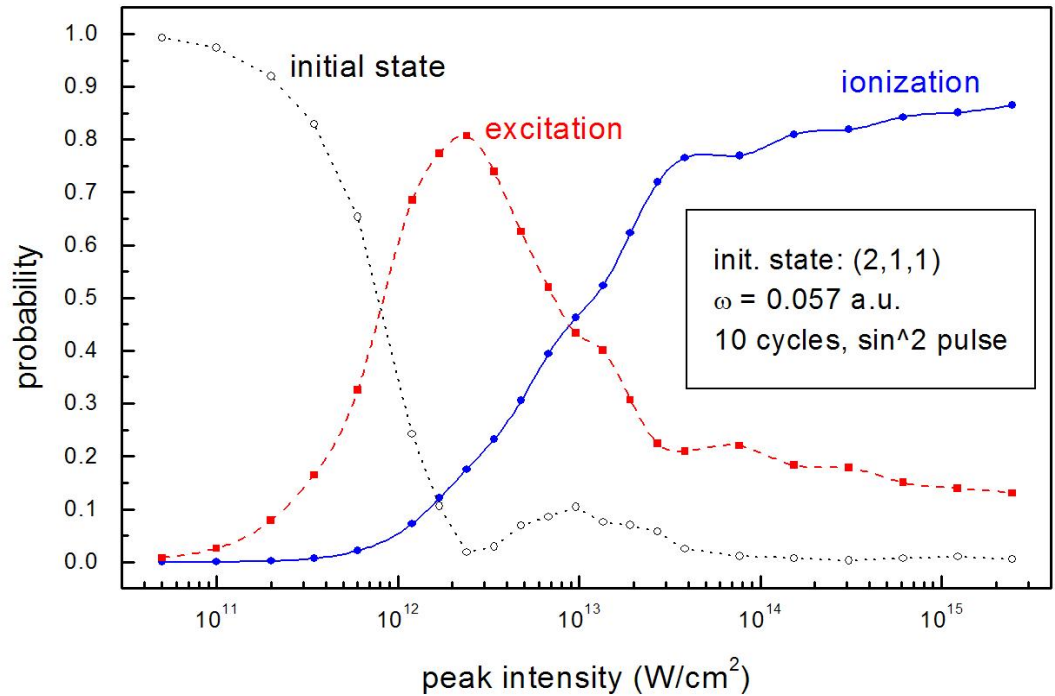

FIG. 2. (Color online) As Fig. 1, but for the initial state $(2,1,1)$. The laser light is co-rotating with respect to the initial state.

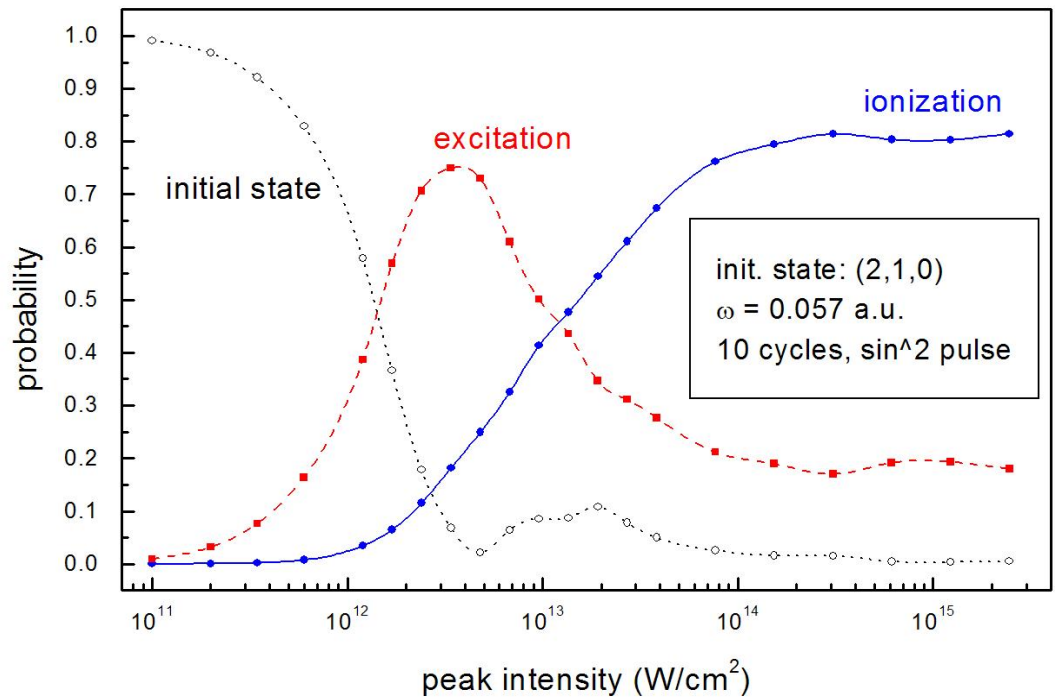

FIG. 3. (Color online) As Fig. 1, but for the initial state $(2,1,0)$.

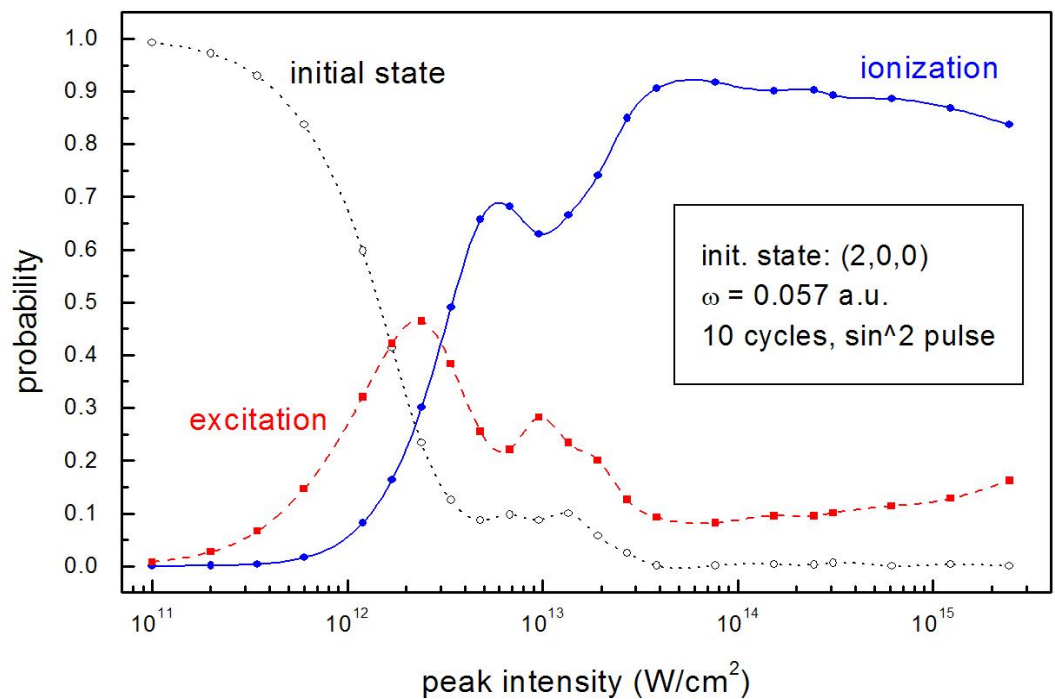

FIG. 4. (Color online) As Fig. 1, but for the initial state $(2,0,0)$.

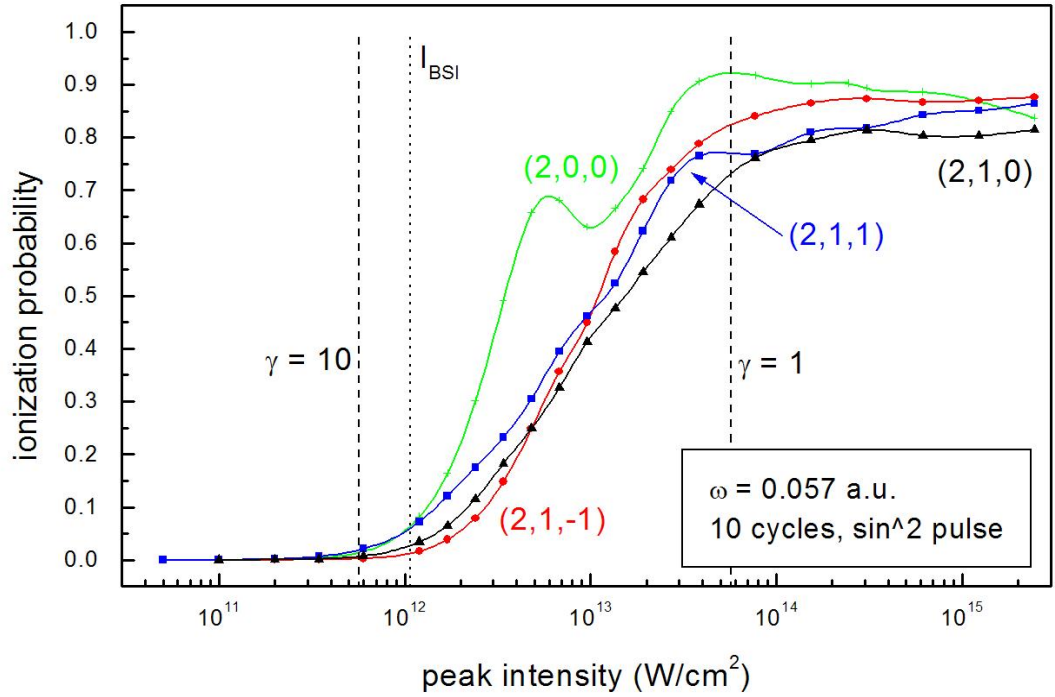

FIG. 5. (Color online) Comparison of ionization probabilities for the four initial states of the hydrogen atom as a function of the peak laser intensity (these lines are identical with respective lines in Figs. 1-4). $(2,1,-1)$ - red line with circles; $(2,1,1)$ - blue line with squares; $(2,1,0)$ - black line with triangles; $(2,0,0)$ - green line with crosses. Two vertical dashed lines correspond to fixed values of the Keldysh parameter $\gamma$, namely $\gamma=10$ and $\gamma=1$. ($\gamma$ decreases from left to right.) The vertical dotted line shows $I_{BSI}=1.1\cdot 10^{12} W/cm^2$.

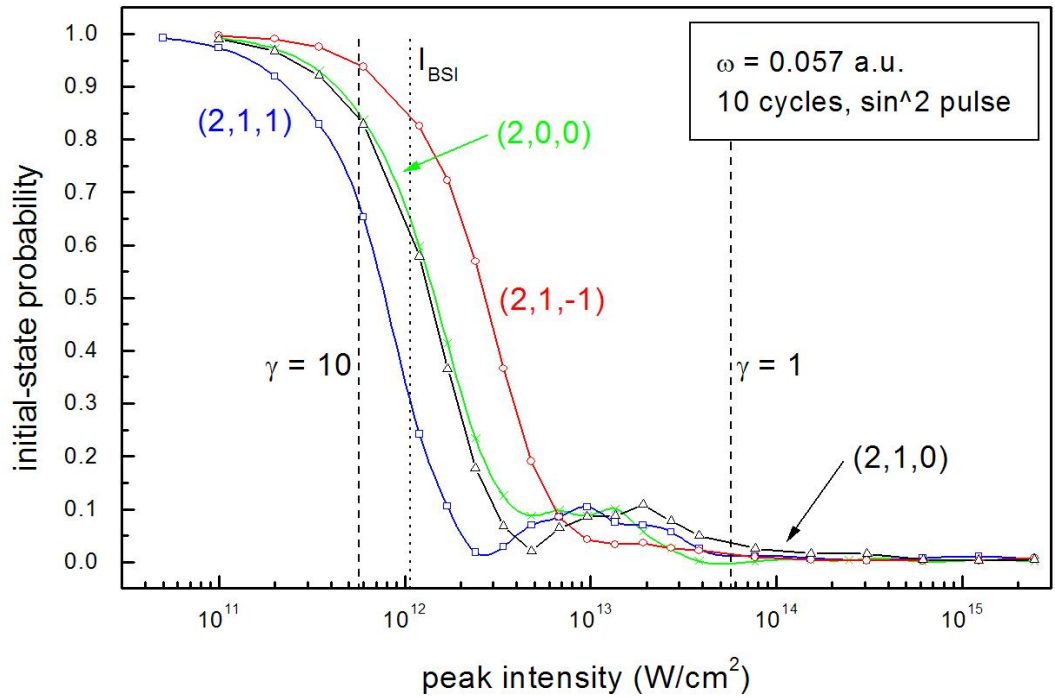

FIG. 6. (Color online) Comparison of initial-state probabilities for the four initial states of the hydrogen atom as a function of the peak laser intensity (these lines are identical with respective lines in Figs. 1-4). $(2,1,-1)$ - red line with circles; $(2,1,1)$ - blue line with squares; $(2,1,0)$ - black line with triangles; $(2,0,0)$ - green line with crosses. Three vertical lines show the same as in Fig. 5.